\documentclass[iop]{emulateapj}
\usepackage{amssymb}
\usepackage{times}
\usepackage{graphicx}
\usepackage{epstopdf}
\usepackage{subfigure}
\usepackage{natbib}
\usepackage{color}

\shorttitle{Halo bias in galaxy clusters}
\shortauthors{Covone et al.}

\newcommand{\jcap}{Journal of Cosmology and Astroparticle Physics}
\newcommand{\kms}{{{\rm \, km~s^{-1}}}}
\newcommand{\bh}{{b_{\rm h}}}  
\newcommand{\beq}{\begin{equation}}
\newcommand{\eeq}{\end{equation}}

\def\ba{\begin{eqnarray}}
\def\ea{\end{eqnarray}}

\begin{document}

\title{Measurement of the halo bias from stacked shear profiles of galaxy clusters}
\author{Giovanni Covone\altaffilmark{1,2}, Mauro Sereno\altaffilmark{3}, Martin Kilbinger\altaffilmark{4}and 
Vincenzo F. Cardone\altaffilmark{5}}
\altaffiltext{1}{Dipartimento di Fisica, Universit\`a di Napoli ''Federico II'', Via Cinthia, I-80126 Napoli, Italy}
\altaffiltext{2}{INFN Sez. di Napoli, Compl. Univ. Monte S. Angelo, Via Cinthia, I-80126 Napoli, Italy}
\altaffiltext{3}{Dipartimento di Fisica e Astronomia, Universit\`a di Bologna, Viale Berti Pichat 6/2, 40127 Bologna, Italia}
\altaffiltext{4}{CEA/Irfu/SAp Saclay, Laboratoire AIM, 91191 Gif-sur-Yvette, France}
\altaffiltext{5}{I.N.A.F. - Osservatorio Astronomico di Roma , Via Frascati 33 - 00040 Monteporzio Catone (Roma), Italy
}

\begin{abstract}

We present the observational evidence of the 2-halo term in the stacked shear profile 
of a sample of $\sim 1200$ optically selected galaxy 
clusters based on imaging data and the public shear catalog from the CFHTLenS.
We find that the halo bias, a measure of the correlated distribution of matter around galaxy clusters, 
has  amplitude and  correlation
with galaxy cluster mass in very good agreement with the predictions based on the LCDM 
standard cosmological model. 
The mass-concentration relation is flat but higher than theoretical predictions.
We also confirm the close scaling relation between the optical richness of galaxy clusters and their mass.

\end{abstract}


\section{Introduction}

A fundamental scope of the observational study of galaxy clusters, the most massive gravitationally bound systems in the cosmos, 
is testing models of structure formation 
and gaining insight into the complex physical interplay between the baryonic and dark components of matter. 
Currently, the most successful framework is based 
on the cosmological constant term and the cold, collisionless dark matter (LCDM), 
which nature is still not understood. 
Numerical simulations have provided clear predictions for the 
structure of dark matter halos (e.g., the so-called mass-concentration relation, \citealt{Bullock2001})
and their clustering properties.

The halo power spectrum $P_{\rm h} (k)$ 
describes the statistical properties of the spatial distribution of massive halos.
Observed structures are in principle differently distributed from the underlying dark matter.
The correlation between the halo density field and the underlying matter distribution
is parametrized by the halo bias parameter, $\bh$.
This is defined as  the ratio of the halo power spectrum to the linear matter \citep{Tinker2010, OguriTakada2011}:
\begin{equation}
b^2_{\rm h} (k)= \frac{P_{\rm h}(k)}{P_{\rm m}(k)} \, . 
\label{eq:bias_def}
\end{equation}
%
Theoretical models of gravitational clustering of dark matter haloes 
provide robust predictions for the bias \citep{Mo1996, Sheth2001}.
It is expected to be constant at large scales and an increasing function of the  peak height.
It is therefore of importance to compare such predictions with observations of galaxy clusters, as a function
of mass and cosmic epoch. 

The correlated matter distribution around galaxy clusters manifests itself as a
2-halo term in the mass density profiles, at scales lager than $\sim 10 \, {\rm Mpc}$,
 its contribution being proportional to $\bh$.
The 2-halo term describes the cumulative effects of the large scale structure 
in which galaxy clusters are located.
While weak gravitational lensing is a powerful tool to measure mass distribution in galaxy clusters 
\citep{KneibNatarajan2011}, the lensing signal expected in the 2-halo term is too low
to be reliably detected in the outskirts of individual systems. 
Stacking the lensing signal of a large number of galaxy clusters enable us to detect at high S/N the
2-halo term. 
Stacked shear profiles of galaxy groups and  clusters have been used 
to study the average properties of radial profiles and the observable scaling relations 
\citep{Sheldon2001, Dahle2003, Okabe2010, Leauthaud2010, Oguri2012MNRAS, Sereno2013, Shan2013}.

\cite{Johnston2007} used  $\sim~1.3\times 10^5$ galaxy groups and clusters from the Sloan Digital Sky Survey 
(SDSS) to obtain stacked lensing signals and infer the relations between  
mass and optical richness and luminosity over a wide range of masses,
and  also to measure the halo bias and  concentration as a function of  mass.
\cite{OguriTakada2011} have proposed a cosmological test based on the 
combination of stacked weak lensing of galaxy clusters with number counts and 
correlation functions, showing that this reduces systematic errors from poorly known redshift distributions and scaling relations.

Several studies have shown observational evidence for the 2-halo term in the stacked galaxy-galaxy lensing profile
\citep{Brimioulle2013, Velander2013}.
In this work, we apply this technique at the galaxy clusters scale.
We use public data from CFHTLenS to measure the lensing signal from 
a sample of optically selected galaxy clusters \citep{Wen2012}.  
As framework we use a spatially flat, cosmological model with $\Omega_\Lambda  = 0.7$, 
$\Omega_{\rm m} = 0.3$ and $h=0.7$, with $H_0 = 100 \, h \, \kms \, {\rm Mpc}^{-1}$.

\section{Analysis}

The  radial shear profile $\gamma(\theta)$ of a galaxy cluster at redshift $z$  can be effectively described
by the sum of two terms:  $\gamma (\theta) = \gamma_{1h} (\theta) + \gamma_{2h}(\theta)$.
The first term $ \gamma_{1h} (\theta) $ accounts  for the main dark matter halo and the associated 
baryonic component. The 2-halo term 
$ \gamma_{2h}(\theta)$ accounts for the correlated distribution of matter  around the cluster.
The tangential shear profile due to the 2-halo term is \citep{OguriTakada2011, OguriHamana2011}
\begin{equation}
\gamma_{t, 2h} (\theta; M, z) = \int \frac{l \, {\rm d}l}{2 \pi} J_2(l \theta) \, 
\frac { \bar{\rho}_m (z) \, \bh (M; z)}{ (1+z)^3 \, \Sigma_{\rm cr} \, D_A^2(z)} \, P_m(k_l; z) \, ,
\label{eq:gamma_t2}
\end{equation}
where $J_2$ is the second order Bessel function, 
$ { \bar{\rho}_m} (z) $ is the mass density at 
$z$,  $P_m(k_l; z)$ the linear power spectrum, $k_l \equiv l / ( (1+z) D_A(z) )$, 
$\Sigma_{\rm cr} $ the lensing critical density,
$D_A (z)$ the angular-diameter distance. 
We used the \cite{Eisenstein1999}'s prescription to calculate the linear matter power spectrum,
and the baryon and neutrino density and scalar spectral index as measured by the \cite{PlanckXVI}.
The uncorrelated matter distribution along the line of sight provides no contribution to the 
stacked shear signal,  therefore the 2-halo term is a function of the correlated matter distribution 
at the same cluster redshift \citep{OguriTakada2011}.

We used the public shear catalog provided by the Canada-France Hawaii Telescope Lensing Survey 
(CFHTLenS, \cite{Heymans2012}) 
containing the photometry, photometric redshifts and ellipticity measurements 
from the four wide fields covering about 154 square degrees. Data have been collected 
within the CFHT Legacy Survey (CFHTLS, \cite{Erben2013}). 
Photometric redshifts were determined from the available optical $ugriz$ observations \citep{Hildebrandt2012},
with accuracy $\simeq 0.04 \, (1+z)$ and a catastrophic outlier rate of about 4\%.

We used the catalog built by \cite{Wen2012} to identify galaxy clusters in the four fields.
The catalog contains positions, photometric redshift, richness, brightest
cluster galaxy (BCG) magnitude for 132,684 optically selected galaxy clusters, up to $z=0.8$.
Galaxy clusters were identified in the SDSS-DR9 imaging data, 
with detection rate $\sim 75\%$ for masses larger than $\sim 0.6 \times 10^{14} \, M_{\odot} $.
The false detection rate is less than 6\% for the whole sample.
The centre of each galaxy cluster is identified with the position of the BCG.
We  selected  galaxy clusters centred in the CFHTLS fields,
with redshift in the range $0.1 \leq z \leq 0.6 $
and at least one radial shear measurement in the inner 2 Mpc/$h$ (corresponding to about 7.5 arcmin at $z=0.3$).
The upper redshift limit was chosen in order to separate robustly the lensing and
background galaxy populations.
Our final sample contains 530, 89,  457, 287 systems (1176 overall) in the four regions, respectively,
with median redshift $z=0.36$.  No further selection is applied. 

For each galaxy cluster, we determined the 
shear profile $\gamma_t (r)$ around its center as briefly outlined hereafter.
The procedure leading to the shape measurements, 
based on the tool {\tt lensfit}, is described in \cite{Miller2013}.
The CFHTLenS catalog provides the raw shear components, $e_1, e_2$. 
These quantities are still affected  by a small multiplicative and additive bias:
we performed the correction
by applying two calibration parameters, $m$ and $c_2$, empirically derived from the data
and provided in the shear catalog \citep{Heymans2012}. 
The calibration reads 
\begin{equation}
e_{{\rm true,} i} = \frac{e_{{\rm m,} i} - c_i}{1 + {\bar m}}  \, \hspace{1cm} (i=1,2) \, ,
\label{eq:calibration}
\end{equation}
where $e_{{\rm m}}$ is the measured ellipticity. 
The average quantity $\bar{m}$ is evaluated in each radial bin, 
taking into account the weight $w$ of the associated shear measurement.

We selected  background lensed sources
behind each  cluster  by using a two colours selection \citep{Medezinski2010, Oguri2012MNRAS}.
We used the following region in the two-colours space: 
\begin{equation}
(g-r < 0.3)      \\ 
\, {\rm OR}     \\
\, (r - i >  1.3)  \\ 
\, {\rm OR }     \\
\, (r - i > g-r) 
\end{equation}
These criteria efficiently select galaxies at $z>0.7$ \citep{Medezinski2010}.
We did not select background sources on the basis of the photometric redshifts, 
as the faint galaxies photometric redshift distribution
shows an artificial peak at $z_{\rm phot}\sim 0.2$ \citep{Hildebrandt2012}.
%
We note that these sources are mostly characterised by a low value of the {\tt odds} parameter,
that quantifies the relative importance of the most likely redshift \citep{Hildebrandt2012},
hinting to possible degeneracies in the redshift determination based only on optical colors.
Finally, we have also excluded all the sources not detected in one of the 3 filters $gri$
or not satisfying the following requirements on the parameters of the CFHTLenS catalog:
ellipticity weight larger than 0, {\tt mask} smaller or equal to 1,
S/N of the {\tt lensfit} measurement larger than 0,  and $i$ magnitude brighter than the local limit magnitude.
The final density of the background galaxies is about 6 galaxies per arcmin$^2$.

We determined the shear profile as a function of the physical length, up 
to about 20 Mpc/$h$ (corresponding to about 2 degrees, at $z=0.3$).
We excluded the central region of angular radius 0.1 Mpc/$h$ 
around the BCG because of the low number of background sources and 
the low accuracy in the determination  of the cluster centre.
The shear profile has been determined in 23 log-spaced radial bins, with 
logarithm spacing $\delta_{\rm log_{10}} = 0.1$.
Finally, in each radial bin, we determined the weighted ellipticity of the 
background sources, hence the tangential and cross component of the shear, $\gamma_t, \, \gamma_x$.

By using photometric redshifts measurements of the lens and source population,
we calculated the excess surface mass density 
\begin{equation}
\Delta \Sigma (R) = \Sigma_{\rm cr} \, g_t(R) \, , 
\end{equation}
where $g_t(R)$ is the reduced tangential shear and then computed the mean in the given annular bin \citep{McKay2001}.

\begin{figure*}[t!]
\centering
\label{fig:profiles}
\caption{Radial profiles of the excess surface mass density $\Delta \Sigma$ for the six samples of galaxy clusters,
binned according to their optical richness $R_{L_*}$. 
Black points are our measurements. The green line is the main galaxy cluster halo, the 
blue line is the contribution from the 2-halo term.
The black line is the overall fitted radial profile. Dashed lines are extrapolation from the best fit model.}
 \begin{tabular}{cc}  
 \includegraphics[width=1.0\textwidth]{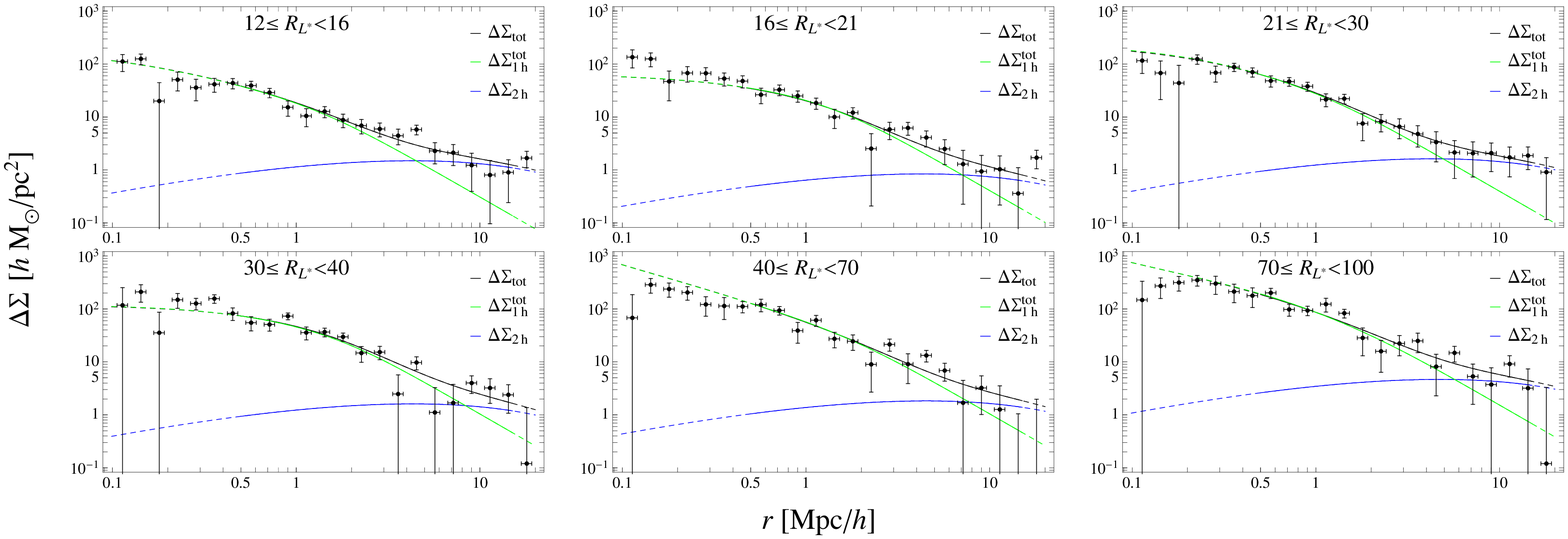}
\end{tabular}
\label{fig:stack}
\end{figure*}

Finally, we have obtained high-S/N radial profiles as a function of the physical length scale 
by stacking the signal of galaxy clusters grouped according to their optical richness.
The optical richness  is defined as $R_{L_*}= L_{200} / L_*$ \citep{Wen2012},
where $ L_{200}$ is the cluster total luminosity within the an empirically determined radius $r_{200}$
and $L_*$ is the evolved characteristic luminosity of galaxies in the $r$ band.
%
%
\cite{Wen2012} have shown that there is a close correlation between the virial mas $M_{200}$
and the optical richness. 
Hence, we used the observable quantity $R_{L_*}$ to split the 
full galaxy cluster sample in $6$ bins, see Table~\ref{table:stacked}.
Stacked radial profiles of the excess mass density $\Delta \Sigma$ are shown in Fig.~\ref{fig:stack}.

We fitted each stacked profile with a double mass component.
The galaxy cluster halo is fitted with the smoothly truncated NFW profile proposed by \cite{Baltz2009JCAP},
\beq
\rho_{\rm BMO} = \frac{\rho_s}{\frac{r}{r_s} (1 + \frac{r}{r_s})^2} \, \left(\frac{r_t^2}{r^2 + r_t^2} \right)^2 \, , 
\eeq
where we set the truncation radius $r_t = 3 \, c_{200} \, r_{s} $.
This modification of the original NFW profile removes the unphysical divergence of its total mass.
Moreover, \cite{OguriHamana2011} used a set of ray-tracing $N$-body simulations to
show that this parametric model gives a less biased estimates of mass and concentration
with respect to the NFW profile, and also better describes the  
density profile beyond the virial radius, where the transition between the cluster and the 2-term halo occurs.
As discussed above, the 2-halo term describes the outer profile.
The shear signal at large radii is proportional to the halo power spectrum, and hence 
to the product $\bh \, \sigma_8^2$, , where $\sigma_8$ is 
 is the rms mass fluctuation amplitude in spheres of size 8$h^{-1}$ Mpc.

A relevant source of error is given by the halo centering offset, as the BCG 
(which defines the cluster center in our sample)
might be misidentified \citep{Johnston2007} or not coincide with the matter centroid \citep{Zitrin2012}.
The BCG offsets are generally small and negligible at the weak lensing scale \citep{Zitrin2012},
but the  mis-identification of the BCG can lead to underestimate
 $\Delta \Sigma(R)$ at small scales and bias low the measurements of $c_{200}$. 
Therefore, we fitted an offset mass component \citep{Yang2006MNRAS373}, 
distributed according to an azimuthally symmetric Gaussian distribution, 
with mass fraction and scale length of the Gaussian distribution as free parameters \citep{Johnston2007}.
The contribution of the main cluster halos is given by the sum of two terms, 
a centred and an off-centred mass component. 
While the effect of miscentering is negligible on the scale where the correlated matter is detected,
it is important to take it into account for an unbiased determination of the cluster concentration. 
We checked that neglecting the offset component leads to underestimate 
$c_{200}$ by $\sim $15\% and the mass by $\sim $10\%. 

Free fitting parameters are the mass and the concentration of the main cluster halo,
the term $\bh \, \sigma_8^2$, the  fraction 
of off-centered halos, $0.5 \leq f_h \leq 1$, 
and the scale length of the
spatial distribution this additional component is extracted from, $0 < \Delta \leq1 \, {\rm Mpc}/h$. 
We performed radial fits using data points  between 0.5 and 15 Mpc/$h$,
beyond which annuli are largely incomplete due to the limited field of view.
The choice of the lower limit for radial range is a compromise between minimising 
the systematic errors due to the contamination of cluster galaxies,
and minimising the statistical error on $c_{200}$ \citep{Mandelbaum2008}.
The number of degrees of freedom  is $10$ in each fit.
Errors were determined with a standard bootstrap procedure with replacement.

In order to determine residual systematic effects affecting the stacked $\gamma_t, \gamma_x$ profiles,
we built a random catalog of 5000 lenses with the same redshift distribution of the real galaxy clusters and
we  determined the lensing signal around the random positions 
using the same procedure described above.
The systematic signal has been then subtracted from the measured quantity. 
Errors on the stacked random clusters shear profiles
have been estimated with a bootstrap procedure with replacement (5000 resamplings)
and then added in quadrature. 
 The stacked lensing signal
$\Delta \Sigma$ from the random pointings is consistent with zero up to $\sim 5~{\rm Mpc}/h$. 
At larger radii, there is a spurious signal at about 1-$\sigma$ level, in agreement 
with \cite{Miyatake2013} who pointed to residual systematics in the shear measurements at the edges of detector.

Due to stacking, shear observations at different radii are correlated. 
The off-diagonal terms of the covariance matrix are very noisy 
\citep{Mandelbaum2013} and can be reliably computed only 
in the most populated bins at small optical richness, where we 
verified that the use of the covariance matrix 
does not significantly change the results. 
For consistency across the different richness bins, we only considered diagonal elements to perform the analysis.


The unbiased profiles were then analyzed using a MCMC procedure to explore the parameter space. 
Results are listed in Table~\ref{table:stacked}.
The mass fraction of the offset component is not  strongly constrained by our data; however, 
it is important to include it the fitting procedure as this leads to less biased measurement of the concentration,
otherwise significatively underestimated. The modelling provides good fits to the data.

\begin{table*}
\centering
\caption{Results of the fit of profiles of the six stacked bins in optical richness. 
Reported central values and uncertainties were obtained as bi-weight estimators of the marginalized probability distributions. 
The $\chi^2$ value refers to the best fit model. The number of d.o.f. is $15-5  = 10$.} 
\begin{tabular}{c c c c c c c }
\hline 
\hline
richness bin& $N_{\rm clus} $ & $R_{L_*} $ & $M_{200}$ & $c_{\rm 200}$ &  $\bh  \sigma_8^2$  & $\chi^2$\\
 & & & ($10^{14} {\rm M}_{\odot}\, h^{-1}$) &   &    \\
12 $ \leq R_{L_*} < $ 16   & 476  & $13.8	\pm 1.1  $  & $ 0.48 \pm 0.09$	& $8.6 \pm 5.8 $   &  $  1.76  \pm 0.39  $  &   11.9 \\
16 $ \leq R_{L_*} < $ 21   &  347 & $18.1	\pm 	1.4 $  & $ 0.51 \pm 0.11$ 	& $4.3 \pm 5.3 $   &   $ 1.20  \pm  0.49   $ &   10.7\\
21 $ \leq R_{L_*} < $ 30   &  216 & $24.7	\pm 2.6$    & $ 0.81 \pm 0.13$ 	& $9.3 \pm 5.5 $   &   $ 1.77    \pm 0.56  $ & 4.45  \\
30 $\leq R_{L_*} < $  40   &  90  & $34.2	\pm 2.7 $   & $ 1.52 \pm 0.24$ 	& $1.8  \pm 1.1 $  &  $ 1.87    \pm  0.94 $  &   18.8 \\
40 $ \leq R_{L_*} < $ 70   &  37  & $47.8  	\pm 6.7$    & $ 1.95 \pm 0.30$ 	& $10.1 \pm 5.6 $ &   $ 2.14   \pm 1.15   $ &   16.2\\
70 $\leq R_{L_*} < $ 100  &  10  &$85.6	\pm 10.3$  & $ 3.21 \pm 0.54$	& $10.4 \pm 5.2 $ &   $ 5.45   \pm 2.31  $ &   14.3\\
\hline
\end{tabular}
\label{table:stacked}
\end{table*}

\section{Scaling relations}

We analyzed scaling relations between clusters observables. 
Linear fits were performed using the BCES method, an implementation of the ordinary least squares estimator \citep{Akritas1996}. 
Errors were estimated by means of a bootstrap resampling with replacements.

\begin{figure}
       \resizebox{\hsize}{!}{\includegraphics{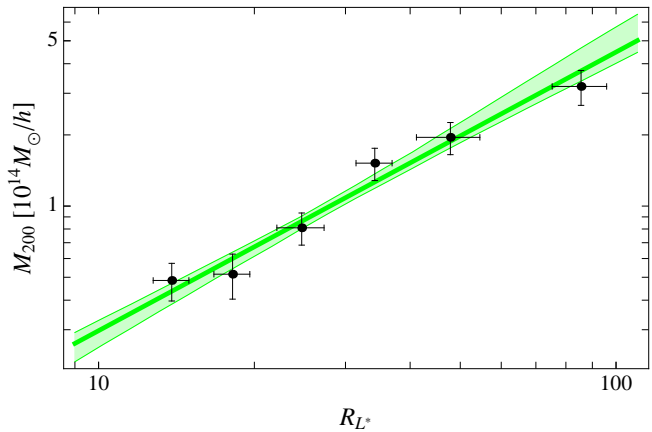}}
       \caption{Correlation between the mass of the galaxy clusters and the optical richness $R_{L_*}$.
       The line and the shaded regions show the  linear relation and its 1-$\sigma$ uncertainty.}
	\label{fig:richness}
\end{figure}

We confirm the strong correlation between the optical richness $R_{L_*}$
and the virial mass of the stacked galaxy clusters, see Fig.~\ref{fig:richness}:
\begin{equation}
{\rm log} \, \frac{M_{200}}{10^{14} M_{\odot} h^{-1}} = 0.035 \pm 0.028 + (1.18 \pm 0.12)  \, {\rm log} \, \frac{R_{L_*}}{ 30} \, ,
\end{equation}
in excellent agreement with \cite{Wen2009}. 

\begin{figure}
       \resizebox{\hsize}{!}{\includegraphics{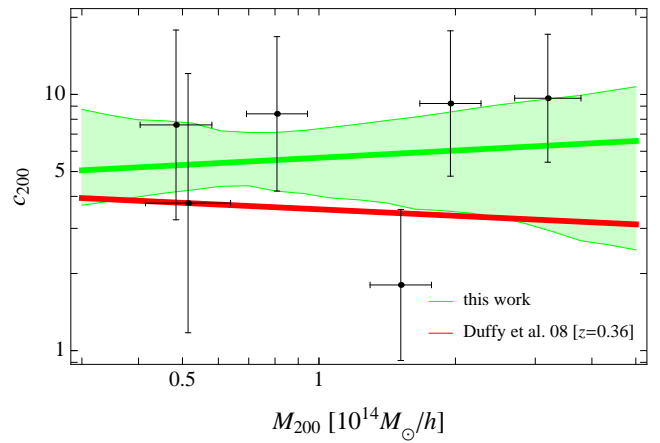}}
       \caption{Concentration-mass relation for the stacked galaxy clusters.
       Green: relation and its 1-$\sigma$ scatter found in this work. Red line: theoretical 
       prediction for individual galaxy clusters by \cite{Duffy2008}.}
	\label{fig:c_M}
\end{figure}

Theoretical models of hierarchical growth of structures in the LCDM model
present two clear predictions on the mass scale of galaxy clusters:
the mass-concentration relation and the dependence of the halo bias term on the parent halo mass.
We find that the stacked galaxy clusters follow a flat $c(M)$-relation, see Fig.~\ref{fig:c_M}: 
\beq
{\rm log} \, c_{200} =  (0.75 \pm 0.13) + (0.09 \pm 0.33) \, {\rm log} \frac{M_{200}}{10^{14} \, M_{\odot} \, h} \, .
\eeq
While the slope is consistent with theoretical predictions by \cite{Duffy2008},
the normalisation differs at 1-$\sigma$ level. 
%
We note that neglecting the offset component would decrease the normalisation, without changing the slope of the relation.

\begin{figure}
       \resizebox{\hsize}{!}{\includegraphics{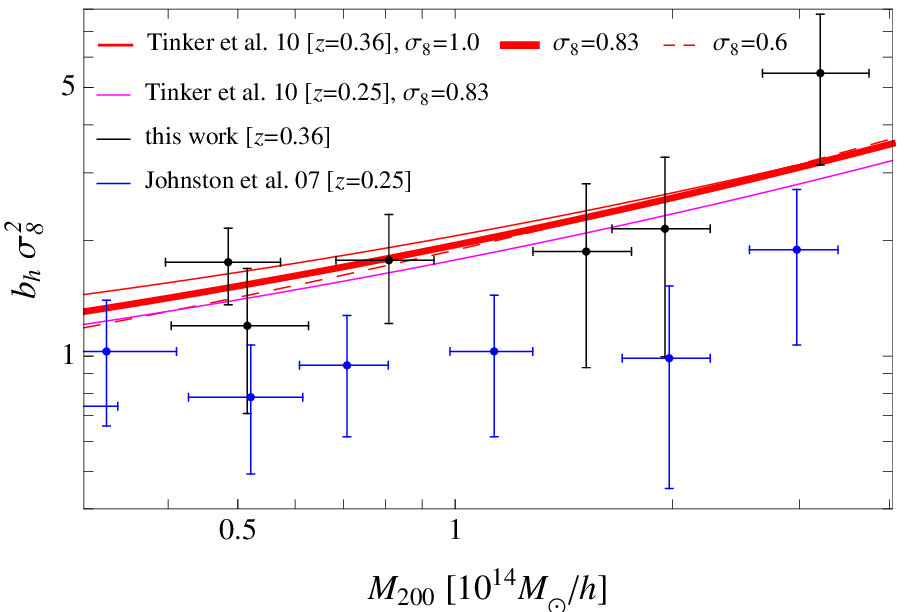}}
       \caption{The quantity $\bh \, \sigma_8^2$ as a function of the halo mass.
       Black points are our measurements  and blue points measurements are from \cite{Johnston2007}. 
Red curves are the theoretical predictions from \cite{Tinker2010} for three fiducial values of  $\sigma_8$.}
\label{fig:bias}
\end{figure}

The 2-halo term is evident in each individual $R_{L_*}$ bin (Fig.~\ref{fig:stack}),
and it is strongly dependent on the parent halo mass (Fig.~\ref{fig:bias}).
Our results  are remarkably consistent with theoretical predictions \citep{Tinker2010,bha+al11}. 
A quantitative analysis can be performed in terms of a $\chi^2$ function,
\beq
\chi^2 = \Sigma_{i=1}^6 \frac{\left(b_{{\rm th},i}\sigma_8^2 - 
[\bh \sigma_8^2]_i\right)^2}{\left( \frac{\partial b_{\rm th}\sigma_8^2}{\partial M_{200}}\right)_i^2 \delta^2_{M_{200}} 
+ \delta^2_{(b_i\sigma_8^2)}} \, .
\label{eq:chi2}
\eeq
We get $\chi^2=2.6$ assuming the theoretical bias from \citet{Tinker2010} and the value $\sigma_8 = 0.83$ 
measured by the Planck collaboration from the CMB 
temperature anisotropy and the lensing-potential power spectra \citep{PlanckXVI}. 
The associated probability is $P(\chi^2 > 2.6) = 86\%$.

In an attempt to constrain $\sigma_8$, we can adopt the bias model from \citet{Tinker2010}. 
We get $\sigma_8 =0.7\pm0.3$.  

\section{Conclusions}

We  used the CFHTLenS public shear catalog to obtain the 
stacked weak lensing signal of 1176 galaxy clusters in the redshift range $0.1 < z < 0.6 \, $.
With respect to previous observational studies of the galaxy cluster halo bias 
based on SDSS imaging data, we used shear measurements
derived from the deeper and higher quality data from the CFHTLS.
We have built high S/N profiles in six bins in optical richness $R_{L_*}$,
which average masses span the range from 0.5 to $\sim 3 \, \times 10^{14} \, M_{\odot}/h$.

We confirm that the cluster mass robustly correlates  with the optical richness,
with parameters in agreement with those found by \cite{Wen2012}.
Our result supports the use of the stacking method in weak lensing to calibrate galaxy 
clusters scaling relations, e.g. \cite{OguriTakada2011} . 

We studied the $c(M)$ and the $\bh (M)$ relations, 
and we found in both cases an agreement with theoretical predictions based on the LCDM cosmological models.
While consistent at 1-$\sigma$ level with the predictions from \cite{Duffy2008},
we find evidence for an over-concentration of the $c(M)$-relation, 
in closer agreement with recent simulations by \cite{Bhattacharya2013}.

Our results partially reconcile present tension between observed $c(M)$ relations and theoretical predictions. 
On one hand, studies of single clusters found very steep and over-concentrated relations 
\citep{Oguri2012MNRAS, fed12, Sereno2013}.
On the other hand, previous stacked analyses found flat relations of the expected amplitude (Mandelbaum et al. 2008). 
We found an alternative scenario on a middling ground: an over-concentrated but still flat relation. 
Analysis of individual clusters might be affected by low S/N and high correlation, 
as well as by selection effects whereas previous SDSS stacked analysis might suffer from contamination and off-set effects.

The large scale shear profile is a degenerate function of the halo bias and the power spectrum normalisation,
therefore our measurement of the halo bias term is not independent of  $\sigma_8$.
Our measurement of the quantity $\bh (M)\, \sigma_8^2$ as a function of the average halo mass is in
very good agreement with theoretical predictions by Tinker et al. (2010),
and systematically higher than the ones by \cite{Johnston2007}, Fig.~\ref{fig:bias}.
As shown by \cite{Rozo2010}, shear measurements in \cite{Johnston2007} 
are underestimated by $\sim$ 20\% due to the diluted lensing signal by foreground galaxies.
This systematic error might explain the discrepancy with our measurements. 

Our analysis is also consistent with measurements from the correlation function of galaxy clusters. 
\citet{Veropalumbo2013} analyzed a spectroscopic sample of 25226 clusters of richness $R_{L_*}\geq 12$ at $z\sim 0.3$ 
and obtained $\bh=2.06\pm0.04$ for $\sigma_8=0.8$.

Our results open the possibility to verify further physical effects on the clustering properties of massive halos: 
\cite{Villaescusa2013} used $N$-body simulations to show that 
cosmological models with massive neutrinos show 
a scale-dependent bias on large scales, while  \cite{Dalal2008} have shown that  
the non-Gaussianity of primordial fluctuations brings to a strongly scale-dependent bias.

\section*{Acknowledgments}
We thank Keiichi Umetsu for stimulating discussions and the referee for clarifying comments.
GC acknowledges financial support from the grant PRIN-INAF 2011 
"Galaxy Evolution with the VLT Survey Telescope";
MS and VFC acknowledge financial support from the agreement ASI/INAF/I/023/12/0.


\begin{thebibliography}{}
\expandafter\ifx\csname natexlab\endcsname\relax\def\natexlab#1{#1}\fi

\bibitem[{{Akritas} \& {Bershady}(1996)}]{Akritas1996}
{Akritas}, M.~G., \& {Bershady}, M.~A. 1996, \apj, 470, 706

\bibitem[{{Baltz} {et~al.}(2009){Baltz}, {Marshall}, \&
  {Oguri}}]{Baltz2009JCAP}
{Baltz}, E.~A., {Marshall}, P., \& {Oguri}, M. 2009, \jcap, 1, 15

\bibitem[{{Bhattacharya} {et~al.}(2013){Bhattacharya}, {Habib}, {Heitmann}, \&
  {Vikhlinin}}]{Bhattacharya2013}
{Bhattacharya}, S., {Habib}, S., {Heitmann}, K., \& {Vikhlinin}, A. 2013, \apj,
  766, 32

\bibitem[{{Bhattacharya} {et~al.}(2011){Bhattacharya}, {Heitmann}, {White},
  {Luki{\'c}}, {Wagner}, \& {Habib}}]{bha+al11}
{Bhattacharya}, S., {Heitmann}, K., {White}, M., {et~al.} 2011, \apj, 732, 122

\bibitem[{{Brimioulle} {et~al.}(2013){Brimioulle}, {Seitz}, {Lerchster},
  {Bender}, \& {Snigula}}]{Brimioulle2013}
{Brimioulle}, F., {Seitz}, S., {Lerchster}, M., {Bender}, R., \& {Snigula}, J.
  2013, \mnras, 432, 1046

\bibitem[{{Bullock} {et~al.}(2001){Bullock}, {Kolatt}, {Sigad}, {Somerville},
  {Kravtsov}, {Klypin}, {Primack}, \& {Dekel}}]{Bullock2001}
{Bullock}, J.~S., {Kolatt}, T.~S., {Sigad}, Y., {et~al.} 2001, \mnras, 321, 559

\bibitem[{{Dahle} {et~al.}(2003){Dahle}, {Hannestad}, \&
  {Sommer-Larsen}}]{Dahle2003}
{Dahle}, H., {Hannestad}, S., \& {Sommer-Larsen}, J. 2003, \apjl, 588, L73

\bibitem[{{Dalal} {et~al.}(2008){Dalal}, {Dor{\'e}}, {Huterer}, \&
  {Shirokov}}]{Dalal2008}
{Dalal}, N., {Dor{\'e}}, O., {Huterer}, D., \& {Shirokov}, A. 2008, \prd, 77,
  123514

\bibitem[{{Duffy} {et~al.}(2008){Duffy}, {Schaye}, {Kay}, \& {Dalla
  Vecchia}}]{Duffy2008}
{Duffy}, A.~R., {Schaye}, J., {Kay}, S.~T., \& {Dalla Vecchia}, C. 2008,
  \mnras, 390, L64

\bibitem[{{Eisenstein} \& {Hu}(1999)}]{Eisenstein1999}
{Eisenstein}, D.~J., \& {Hu}, W. 1999, \apj, 511, 5

\bibitem[{{Erben} {et~al.}(2013){Erben}, {Hildebrandt}, {Miller}, {van
  Waerbeke}, {Heymans}, {Hoekstra}, {Kitching}, {Mellier}, {Benjamin}, {Blake},
  {Bonnett}, {Cordes}, {Coupon}, {Fu}, {Gavazzi}, {Gillis}, {Grocutt}, {Gwyn},
  {Holhjem}, {Hudson}, {Kilbinger}, {Kuijken}, {Milkeraitis}, {Rowe},
  {Schrabback}, {Semboloni}, {Simon}, {Smit}, {Toader}, {Vafaei}, {van Uitert},
  \& {Velander}}]{Erben2013}
{Erben}, T., {Hildebrandt}, H., {Miller}, L., {et~al.} 2013, \mnras, 433, 2545

\bibitem[{{Fedeli}(2012)}]{fed12}
{Fedeli}, C. 2012, \mnras, 424, 1244

\bibitem[{{Heymans} {et~al.}(2012){Heymans}, {Van Waerbeke}, {Miller}, {Erben},
  {Hildebrandt}, {Hoekstra}, {Kitching}, {Mellier}, {Simon}, {Bonnett},
  {Coupon}, {Fu}, {Harnois D{\'e}raps}, {Hudson}, {Kilbinger}, {Kuijken},
  {Rowe}, {Schrabback}, {Semboloni}, {van Uitert}, {Vafaei}, \&
  {Velander}}]{Heymans2012}
{Heymans}, C., {Van Waerbeke}, L., {Miller}, L., {et~al.} 2012, \mnras, 427,
  146

\bibitem[{{Hildebrandt} {et~al.}(2012){Hildebrandt}, {Erben}, {Kuijken}, {van
  Waerbeke}, {Heymans}, {Coupon}, {Benjamin}, {Bonnett}, {Fu}, {Hoekstra},
  {Kitching}, {Mellier}, {Miller}, {Velander}, {Hudson}, {Rowe}, {Schrabback},
  {Semboloni}, \& {Ben{\'{\i}}tez}}]{Hildebrandt2012}
{Hildebrandt}, H., {Erben}, T., {Kuijken}, K., {et~al.} 2012, \mnras, 421, 2355

\bibitem[{{Johnston} {et~al.}(2007){Johnston}, {Sheldon}, {Wechsler}, {Rozo},
  {Koester}, {Frieman}, {McKay}, {Evrard}, {Becker}, \& {Annis}}]{Johnston2007}
{Johnston}, D.~E., {Sheldon}, E.~S., {Wechsler}, R.~H., {et~al.} 2007, ArXiv
  e-prints, arXiv:0709.1159

\bibitem[{{Kneib} \& {Natarajan}(2011)}]{KneibNatarajan2011}
{Kneib}, J.-P., \& {Natarajan}, P. 2011, \aapr, 19, 47

\bibitem[{{Leauthaud} {et~al.}(2010){Leauthaud}, {Finoguenov}, {Kneib},
  {Taylor}, {Massey}, {Rhodes}, {Ilbert}, {Bundy}, {Tinker}, {George}, {Capak},
  {Koekemoer}, {Johnston}, {Zhang}, {Cappelluti}, {Ellis}, {Elvis}, {Giodini},
  {Heymans}, {Le F{\`e}vre}, {Lilly}, {McCracken}, {Mellier},
  {R{\'e}fr{\'e}gier}, {Salvato}, {Scoville}, {Smoot}, {Tanaka}, {Van
  Waerbeke}, \& {Wolk}}]{Leauthaud2010}
{Leauthaud}, A., {Finoguenov}, A., {Kneib}, J.-P., {et~al.} 2010, \apj, 709, 97

\bibitem[{{Mandelbaum} {et~al.}(2008){Mandelbaum}, {Seljak}, \&
  {Hirata}}]{Mandelbaum2008}
{Mandelbaum}, R., {Seljak}, U., \& {Hirata}, C.~M. 2008, \jcap, 8, 6

\bibitem[{{Mandelbaum} {et~al.}(2013){Mandelbaum}, {Slosar}, {Baldauf},
  {Seljak}, {Hirata}, {Nakajima}, {Reyes}, \& {Smith}}]{Mandelbaum2013}
{Mandelbaum}, R., {Slosar}, A., {Baldauf}, T., {et~al.} 2013, \mnras, 432, 1544

\bibitem[{{McKay} {et~al.}(2001){McKay}, {Sheldon}, {Racusin}, {Fischer},
  {Seljak}, {Stebbins}, {Johnston}, {Frieman}, {Bahcall}, {Brinkmann},
  {Csabai}, {Fukugita}, {Hennessy}, {Ivezic}, {Lamb}, {Loveday}, {Lupton},
  {Munn}, {Nichol}, {Pier}, \& {York}}]{McKay2001}
{McKay}, T.~A., {Sheldon}, E.~S., {Racusin}, J., {et~al.} 2001, ArXiv
  Astrophysics e-prints, astro-ph/0108013

\bibitem[{{Medezinski} {et~al.}(2010){Medezinski}, {Broadhurst}, {Umetsu},
  {Oguri}, {Rephaeli}, \& {Ben{\'{\i}}tez}}]{Medezinski2010}
{Medezinski}, E., {Broadhurst}, T., {Umetsu}, K., {et~al.} 2010, \mnras, 405,
  257

\bibitem[{{Miller} {et~al.}(2013){Miller}, {Heymans}, {Kitching}, {van
  Waerbeke}, {Erben}, {Hildebrandt}, {Hoekstra}, {Mellier}, {Rowe}, {Coupon},
  {Dietrich}, {Fu}, {Harnois-D{\'e}raps}, {Hudson}, {Kilbinger}, {Kuijken},
  {Schrabback}, {Semboloni}, {Vafaei}, \& {Velander}}]{Miller2013}
{Miller}, L., {Heymans}, C., {Kitching}, T.~D., {et~al.} 2013, \mnras, 429,
  2858

\bibitem[{{Miyatake} {et~al.}(2013){Miyatake}, {More}, {Mandelbaum}, {Takada},
  {Spergel}, {Kneib}, {Schneider}, {Brinkmann}, \& {Brownstein}}]{Miyatake2013}
{Miyatake}, H., {More}, S., {Mandelbaum}, R., {et~al.} 2013, ArXiv e-prints,
  arXiv:1311.1480

\bibitem[{{Mo} \& {White}(1996)}]{Mo1996}
{Mo}, H.~J., \& {White}, S.~D.~M. 1996, \mnras, 282, 347

\bibitem[{{Oguri} {et~al.}(2012){Oguri}, {Bayliss}, {Dahle}, {Sharon},
  {Gladders}, {Natarajan}, {Hennawi}, \& {Koester}}]{Oguri2012MNRAS}
{Oguri}, M., {Bayliss}, M.~B., {Dahle}, H., {et~al.} 2012, \mnras, 420, 3213

\bibitem[{{Oguri} \& {Hamana}(2011)}]{OguriHamana2011}
{Oguri}, M., \& {Hamana}, T. 2011, \mnras, 414, 1851

\bibitem[{{Oguri} \& {Takada}(2011)}]{OguriTakada2011}
{Oguri}, M., \& {Takada}, M. 2011, \prd, 83, 023008

\bibitem[{{Okabe} {et~al.}(2010){Okabe}, {Takada}, {Umetsu}, {Futamase}, \&
  {Smith}}]{Okabe2010}
{Okabe}, N., {Takada}, M., {Umetsu}, K., {Futamase}, T., \& {Smith}, G.~P.
  2010, \pasj, 62, 811

\bibitem[{{Planck Collaboration} {et~al.}(2013){Planck Collaboration}, {Ade},
  {Aghanim}, {Armitage-Caplan}, {Arnaud}, {Ashdown}, {Atrio-Barandela},
  {Aumont}, {Baccigalupi}, {Banday}, \& et~al.}]{PlanckXVI}
{Planck Collaboration}, {Ade}, P.~A.~R., {Aghanim}, N., {et~al.} 2013, ArXiv
  e-prints, arXiv:1303.5076

\bibitem[{{Rozo} {et~al.}(2010){Rozo}, {Wechsler}, {Rykoff}, {Annis}, {Becker},
  {Evrard}, {Frieman}, {Hansen}, {Hao}, {Johnston}, {Koester}, {McKay},
  {Sheldon}, \& {Weinberg}}]{Rozo2010}
{Rozo}, E., {Wechsler}, R.~H., {Rykoff}, E.~S., {et~al.} 2010, \apj, 708, 645

\bibitem[{{Sereno} \& {Covone}(2013)}]{Sereno2013}
{Sereno}, M., \& {Covone}, G. 2013, \mnras, 434, 878

\bibitem[{{Shan} {et~al.}(2013){Shan}, {Kneib}, {Comparat}, {Jullo},
  {Charbonnier}, {Erben}, {Makler}, {Moraes}, {Van Waerbeke}, {Courbin},
  {Meylan}, {Tao}, \& {Taylor}}]{Shan2013}
{Shan}, H., {Kneib}, J.-P., {Comparat}, J., {et~al.} 2013, ArXiv e-prints,
  arXiv:1311.1319

\bibitem[{{Sheldon} {et~al.}(2001){Sheldon}, {Annis}, {B{\"o}hringer},
  {Fischer}, {Frieman}, {Joffre}, {Johnston}, {McKay}, {Miller}, {Nichol},
  {Stebbins}, {Voges}, {Anderson}, {Bahcall}, {Brinkmann}, {Brunner}, {Csabai},
  {Fukugita}, {Hennessy}, {Ivezi{\'c}}, {Lupton}, {Munn}, {Pier}, \&
  {York}}]{Sheldon2001}
{Sheldon}, E.~S., {Annis}, J., {B{\"o}hringer}, H., {et~al.} 2001, \apj, 554,
  881

\bibitem[{{Sheth} {et~al.}(2001){Sheth}, {Mo}, \& {Tormen}}]{Sheth2001}
{Sheth}, R.~K., {Mo}, H.~J., \& {Tormen}, G. 2001, \mnras, 323, 1

\bibitem[{{Tinker} {et~al.}(2010){Tinker}, {Robertson}, {Kravtsov}, {Klypin},
  {Warren}, {Yepes}, \& {Gottl{\"o}ber}}]{Tinker2010}
{Tinker}, J.~L., {Robertson}, B.~E., {Kravtsov}, A.~V., {et~al.} 2010, \apj,
  724, 878

\bibitem[{{Velander} {et~al.}(2014){Velander}, {van Uitert}, {Hoekstra},
  {Coupon}, {Erben}, {Heymans}, {Hildebrandt}, {Kitching}, {Mellier}, {Miller},
  {Van Waerbeke}, {Bonnett}, {Fu}, {Giodini}, {Hudson}, {Kuijken}, {Rowe},
  {Schrabback}, \& {Semboloni}}]{Velander2013}
{Velander}, M., {van Uitert}, E., {Hoekstra}, H., {et~al.} 2014, \mnras, 437,
  2111

\bibitem[{{Veropalumbo} {et~al.}(2013){Veropalumbo}, {Marulli}, {Moscardini},
  {Moresco}, \& {Cimatti}}]{Veropalumbo2013}
{Veropalumbo}, A., {Marulli}, F., {Moscardini}, L., {Moresco}, M., \&
  {Cimatti}, A. 2013, ArXiv e-prints, arXiv:1311.5895

\bibitem[{{Villaescusa-Navarro} {et~al.}(2013){Villaescusa-Navarro}, {Marulli},
  {Viel}, {Branchini}, {Castorina}, {Sefusatti}, \& {Saito}}]{Villaescusa2013}
{Villaescusa-Navarro}, F., {Marulli}, F., {Viel}, M., {et~al.} 2013, ArXiv
  e-prints, arXiv:1311.0866

\bibitem[{{Wen} {et~al.}(2009){Wen}, {Han}, \& {Liu}}]{Wen2009}
{Wen}, Z.~L., {Han}, J.~L., \& {Liu}, F.~S. 2009, \apjs, 183, 197

\bibitem[{{Wen} {et~al.}(2012){Wen}, {Han}, \& {Liu}}]{Wen2012}
---. 2012, \apjs, 199, 34

\bibitem[{{Yang} {et~al.}(2006){Yang}, {Mo}, {van den Bosch}, {Jing},
  {Weinmann}, \& {Meneghetti}}]{Yang2006MNRAS373}
{Yang}, X., {Mo}, H.~J., {van den Bosch}, F.~C., {et~al.} 2006, \mnras, 373,
  1159

\bibitem[{{Zitrin} {et~al.}(2012){Zitrin}, {Bartelmann}, {Umetsu}, {Oguri}, \&
  {Broadhurst}}]{Zitrin2012}
{Zitrin}, A., {Bartelmann}, M., {Umetsu}, K., {Oguri}, M., \& {Broadhurst}, T.
  2012, \mnras, 426, 2944

\end{thebibliography}

\end{document}